\documentclass[12pt]{article}
\usepackage[utf8]{inputenc}
\usepackage{graphicx}
\usepackage{amsmath}

\begin{document}


	\begin{center}
	
	{\Large{The Relativistic one dimensional Coulomb problem and Relativistic PIMC method  }}\\[9pt]		
	
	{\large A. Ivanov$^{1 a}$, O. Pavlovsky$^{1,2 b}$}\\[6pt]
	
	\parbox{.9\textwidth}{\centering\small\it
		$^1$ Faculty of Physics, Moscow State University\\
		119991, Moscow, Russia\\
		$^2$ Institute for Theoretical and Experimental Physics,\\
		117218, Moscow, Russia\\
		E-mail: $^a$ivanov.as@physics.msu.ru  $^b$pavlovsky@physics.msu.ru}\\[1cc]
\end{center}

{\parindent5mm
	
{\footnotesize The Relativistic one dimensional Coulomb problem was studied by means of the Path Integral Monte Carlo  method. Relativistic and non-relativistic regimes of this problem were investigated. The relativistic regime appears at small masses of the particle and (or) at strong coupling. Critical coupling at which the bound state disappears was founded. This law is valid for any finite regularization of the Coulomb potential and potentially can be tested experimentally.  }


\section{Introduction}

One dimensional Coulomb problem attracts a lot of interest among the scientists during a long time.  This problem is an important one for many reasons: as a pedagogical problem; as a reduction of multi-dimensional tasks;  as an effective model for study of one dimensional materials like quantum wires or carbon nanotubes. 

The first physical application of the 1D Coulomb problem was proposed in \cite{Loundon_first} as the model for studying the electron properties of hydrogen atom in the strong magnetic field. Today we know another important applications of 1D Coulomb problem in physics of   quantum wires and carbon nanotubes \cite{Loundon2}.  

The aim of our work is to study the relativistic effects in the 1D Coulomb problem. Why these relativistic effects are interesting? The relativistic effects in quantum theory  are associated not only with high particle velocities, but also with strong external fields. Another reason for the importance of relativistic effects is  new materials with  massless electron excitations  \cite{graphene}, \cite{tube}, \cite{semimetals}.   In these cases, the dispersion relations contain Dirac cones and so the effective theory for such quasiparticles  must be quasirelativistic one.  The Coulomb problem in these materials must have  relativistic form. Finally, it is important to note that relativistic Coulomb problem has significant theoretical value too. It is very important to know the limits of applicability of the non-relativistic approximation in the Coulomb problem. 

Non-relativistic 1D Coulomb problem can be studied analytically \cite{anal}, \cite{anal2}, \cite{anal3}, \cite{anal4},  \cite{anal5}, \cite{anal6},  on this way there were obtained many interesting results. Unfortunately, analytical  study of relativistic Coulomb problem is much more difficult task, the Dirac or Klein-Gordon  equations must be used \cite{relanal}, \cite{relanal2}. In our work the Relativistic Path Integral Monte Carlo method was used for studying the properties of the 1D relativistic particle  in the regularized  Coulomb potential.  Path integral Monte Carlo (PIMC) method is one of the most popular numerical approach to the quantum models study \cite{ceperley}. This method becomes especially useful in properties modelling of quantum many-body problems. Relativistic generalisation of this method was proposed in \cite{inp}, where this approach was tested on the problem of quantum relativistic oscillator. The main feature of this study in comparison with the results of \cite{inp} is that in this case we deal with singular potential and some regularization procedure must be proposed for correct definition of Path Integral \cite{kl}, \cite{zeen}.  In relativistic Path Integral formalism the situation with singular potentials is the similar with non-relativistic one. At any finite value of times step $a_t$ we must consider the regularized  Coulomb potential to avoid falling of the trajectories on the singularity point.  For any finite value of time step  $a_t$  there is a critical value of regularization parameter when the Monte Carlo trajectories fall on singularity  and Monte Carlo method gives us incorrect result.  In order to solve this problem the regularization of the singular potential must be  synchronized with taking a continuous limit $a_t \to 0$ \cite{kl}. Fortunately, in the many applications there are some physical reasons for regularization of Coulomb potential singularity. Particularly in the case of the hydrogen atom in strong magnetic field the potential looks as following

\begin{equation}
    V(q, a)= -\frac{\alpha e^2}{\sqrt{q ^ 2 + a ^ 2}},
    \label{pots}
\end{equation}

where $a$ is a regularization parameter \cite{Loundon2}. The same situation occurs in graphene where strong Coulomb interaction between quasiparticles are screened by $\sigma$-orbitals \cite{wehling} and interaction potential can be fit by the function (\ref{pots})  \cite{sigmaorb}. 

In our work we interact with  the phenomenon of the generation of the bound states in the regularized Coulomb potential $V(q, a)$. We will show by means of the Relativistic PIMC method that for any finite value of the regularization parameter $a$ there is a critical value of coupling $\alpha$ at which the bound state disappears and only free motion of the relativistic particle is possible. This phenomenon looks very promising for the experimental tests.  

\section{Relativistic Path Integral Monte-Carlo method}
The Hamiltonian function of the relativistic quantum mechanical system with instantaneous interaction is
\begin{equation}
    H = \sqrt{{\bf p} ^ 2 + m ^ 2} + V ({\bf q}).
\end{equation}
The corresponding density matrix for this system is the following function (see $\cite{inp}$)
\begin{equation*}
    \rho ({\bf{q ''}}, {\bf{q '}}; \tau) = \bigg ( \frac{m \tau}{\pi \sqrt{\tau ^ 2 + ({\bf{q ''}} - {\bf{q '}}) ^ 2}} \bigg ) ^ {(d + 1) / 2} \times
\end{equation*}
\begin{equation}
    \times \frac{K _ {(d + 1) / 2} \bigl ( m \sqrt{\tau ^ 2 + ({\bf{q ''}} - {\bf{q '}}) ^ 2} \bigr )}{(2 \tau) ^ {(d - 1) / 2}} e ^ {-\tau V ({\bf q '})},
\end{equation}
where ${\bf q '}$ --- initial point and ${\bf q ''}$ --- final point, $\tau$ is a time interval and $d$ --- number of degrees of freedom.

Let us consider one dimensional system on the lattice, corresponding average values of the kinetic (see $\cite{inp}$) and potential energies are following
\begin{equation}
	\langle \sqrt{p ^ 2 + m ^ 2} \rangle = \bigg \langle \frac{m \tau}{\sqrt{\tau ^ 2 + (\Delta q) ^ 2}} \frac{K _ 0 (m \sqrt{\tau ^ 2 + (\Delta q) ^ 2})}{K _ 1 (m \sqrt{\tau ^ 2 + (\Delta q) ^ 2})} + \frac{\tau ^ 2 - (\Delta q) ^ 2}{\tau (\tau ^ 2 + (\Delta q) ^ 2)} \bigg \rangle
\end{equation}
\begin{equation}
    \langle V (q) \rangle = \langle V (q _ i) \rangle
\end{equation}

\section{Relativistic Coulomb problem}

Let us consider one dimensional Relativistic Coulomb problem with hamiltonian function

\begin{equation}
    H = \sqrt{{\bf p} ^ 2 + m ^ 2} - \frac{\alpha e^2}{\sqrt{{\bf q} ^ 2 + a ^ 2}}.
\end{equation}

In our consideration  this problem will be studied by means of Relativistic Path Integral Monte-Carlo method. The existence of a bound state in the one-dimensional Coulomb potential is an interesting physical problem and we will study this task numerically.  

Another interesting question is about the relativistic regime in this problem. This regime connects with not only small mass of the particle $m$ but with the strong potential $V$ too.  

In order to find the criterion of the relativistic regime and to test of our numerical approach, the Virial Theorem will be used. 

\subsection{The Virial Theorem}

According to the classical approach, one can write virial theorem for the non-relativistic (NR) one dimensional Coulomb hamiltonian function
\begin{equation}
    H _ {NR} = m + K (p) + V (q),
\end{equation}
where $K (p)$ is operator of kinetic energy
\begin{equation}
    K (p) = \frac{p ^ 2}{2 m}
\end{equation}
and $V (q)$ is operator of potential energy
\begin{equation}
    V (q) = - \frac{\alpha e ^ 2}{\sqrt{q ^ 2 + a ^ 2}}.
\end{equation}
The virial theorem for this system is
\begin{equation}
    2 \langle K \rangle = \langle q \cdot \frac{d V}{d q} \rangle
\end{equation}
or
\begin{equation}
    2 \langle K \rangle = - \langle V + D \rangle,
\end{equation}
where $D =  a ^ 2 \alpha e ^ 2 / (q ^ 2 + a ^ 2) ^ {3 / 2}$ is a correction, which occurs due to regularization parameter $a$.

In next section, we will compere this relation between kinetic and potential energy with our numerical results.    

\newpage
\section{Numerical results}

\subsection{Numerical test of the virial theorem}

The relation between kinetic and potential energies in non-relativistic regime is
\begin{equation}
    2 \langle K \rangle = - \langle V + D \rangle,
\end{equation}
where $D =  a ^ 2 \alpha e ^ 2 / (q ^ 2 + a ^ 2) ^ {3 / 2}$.

The limit $a \to 0$ sets $D \to 0$ and we have classical virial theorem, so the existence of $D$ is caused by lattice approximation of the system.
\begin{equation}
    2 \langle K \rangle = - \langle V \rangle
\end{equation}

Figure $\ref{VirialTheorem}$.b shows that in the case of huge mass $m = 100$ there is a good agreement with relation
\begin{equation}
    2 \langle K \rangle = - \langle V + D \rangle
\end{equation}
in the range of bound state conditions excepting the case of permanent falling of the particle on the attractive center.
The case of small masses $m = 0.1$ has no agreement with non-relativistic virial theorem (see Figure 1.a).
\begin{figure}[ht]
\begin{minipage}[h]{0.48\linewidth}
\center{\includegraphics[width=1.1\linewidth]{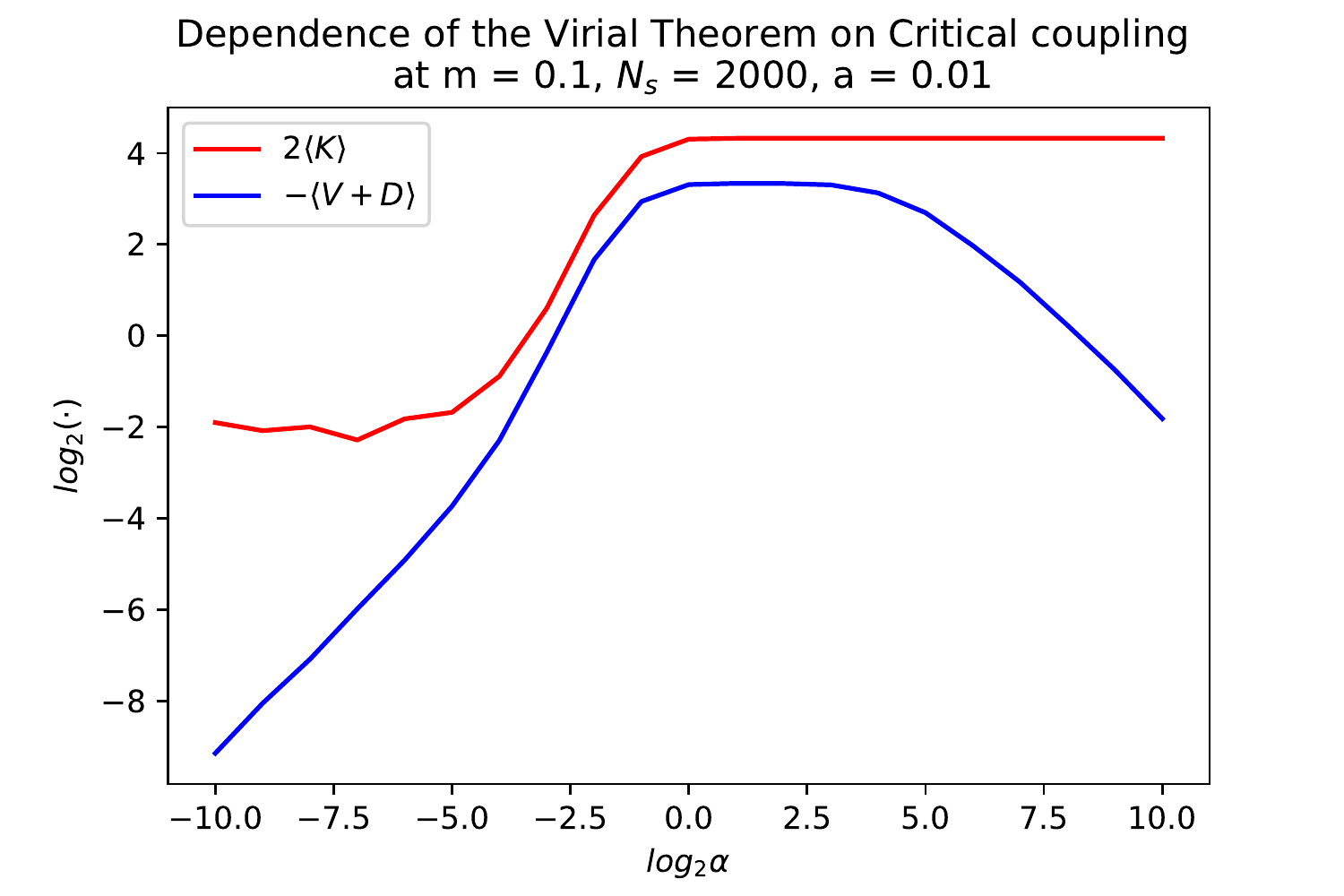} \\ a) $m = 0.1$}
\end{minipage}
\hfill
\begin{minipage}[ht]{0.48\linewidth}
\center{\includegraphics[width=1.1\linewidth]{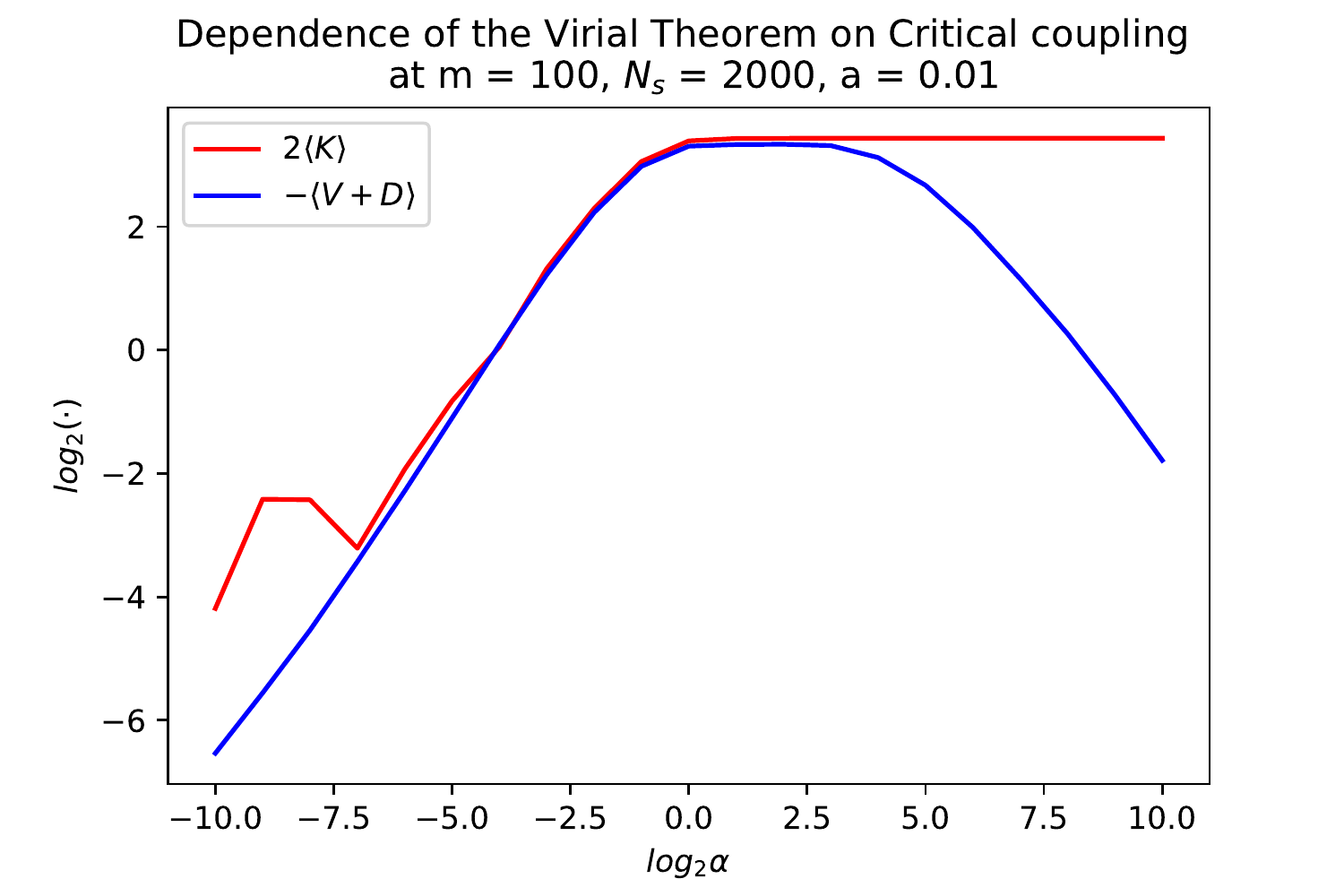} \\ b) $m = 100$}
\end{minipage}
\caption{The virial theorem at $N _ s = 2000$, $a = 0.01$ and different masses}
\label{VirialTheorem}
\end{figure}

\subsection{Existence of the bound state}
By decreasing $\alpha$ one can see the phase transition from bound state to free motion of the particle, where it is no more caught by the attractive center.
There is a critical coupling value $\alpha _ {cr} \approx 2 ^ {-1}$, which separates the states at $m = 1$ and $a = 0.1$ (see Figure 2). Values of coupling constant which are bigger then critical coupling constant corresponds to thermalized condition, which is not depended on number of sweeps.
If the value of coupling constant is less then critical value, the dependence of $\langle q ^ 2 \rangle$ on $N _ s$ is similar to random walk (see Figure 3).
\begin{figure}[ht]
    \centering
    \includegraphics[width=0.75\linewidth]{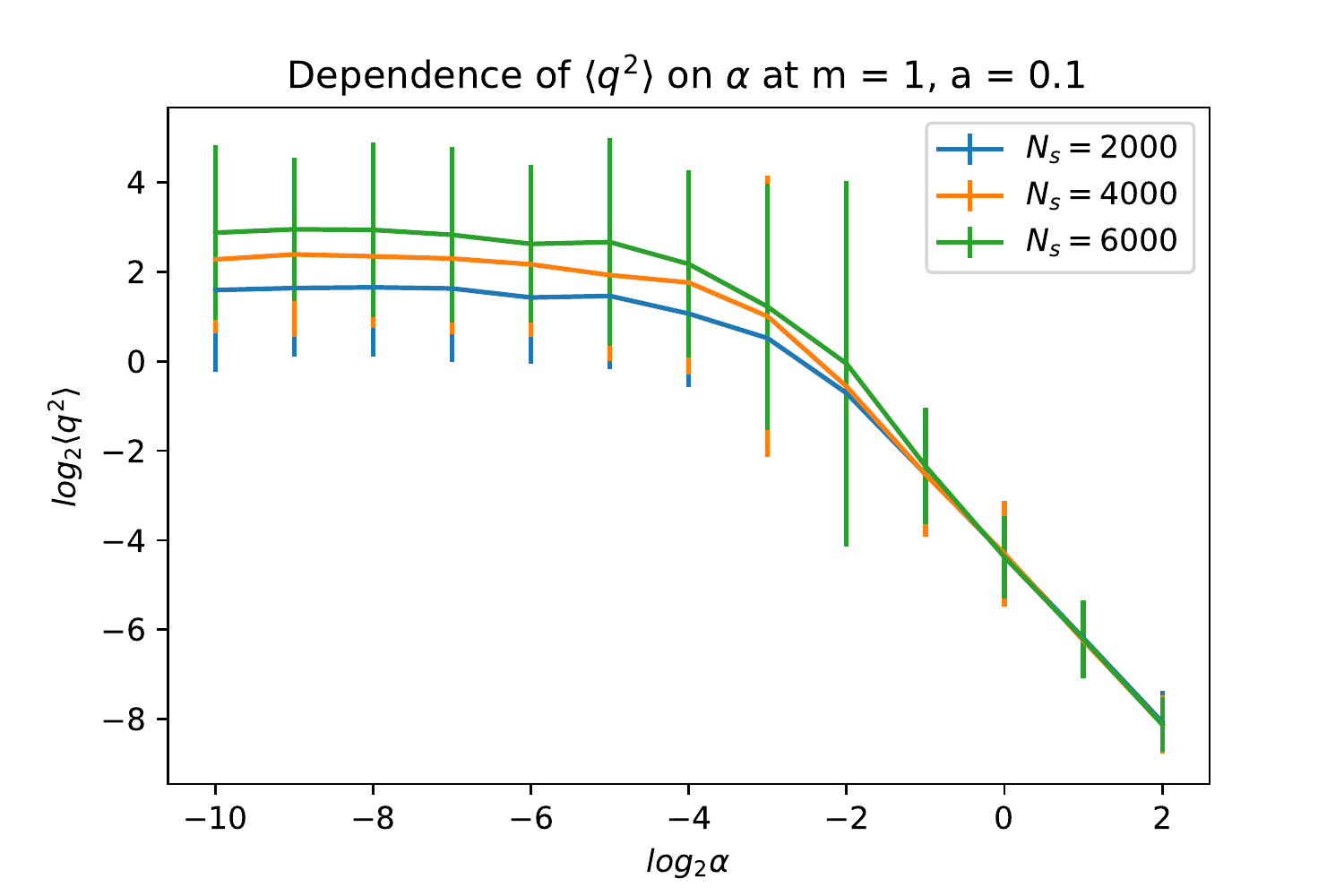}
    \caption{Dependence of $\log_ 2 \langle q ^ 2 \rangle$ on $\log_ 2 \alpha$ at $m = 1$, $a = 0.01$ and different values of $N _ s$.}
\end{figure}

\begin{figure}[ht]
    \centering
    \includegraphics[width=0.75\linewidth]{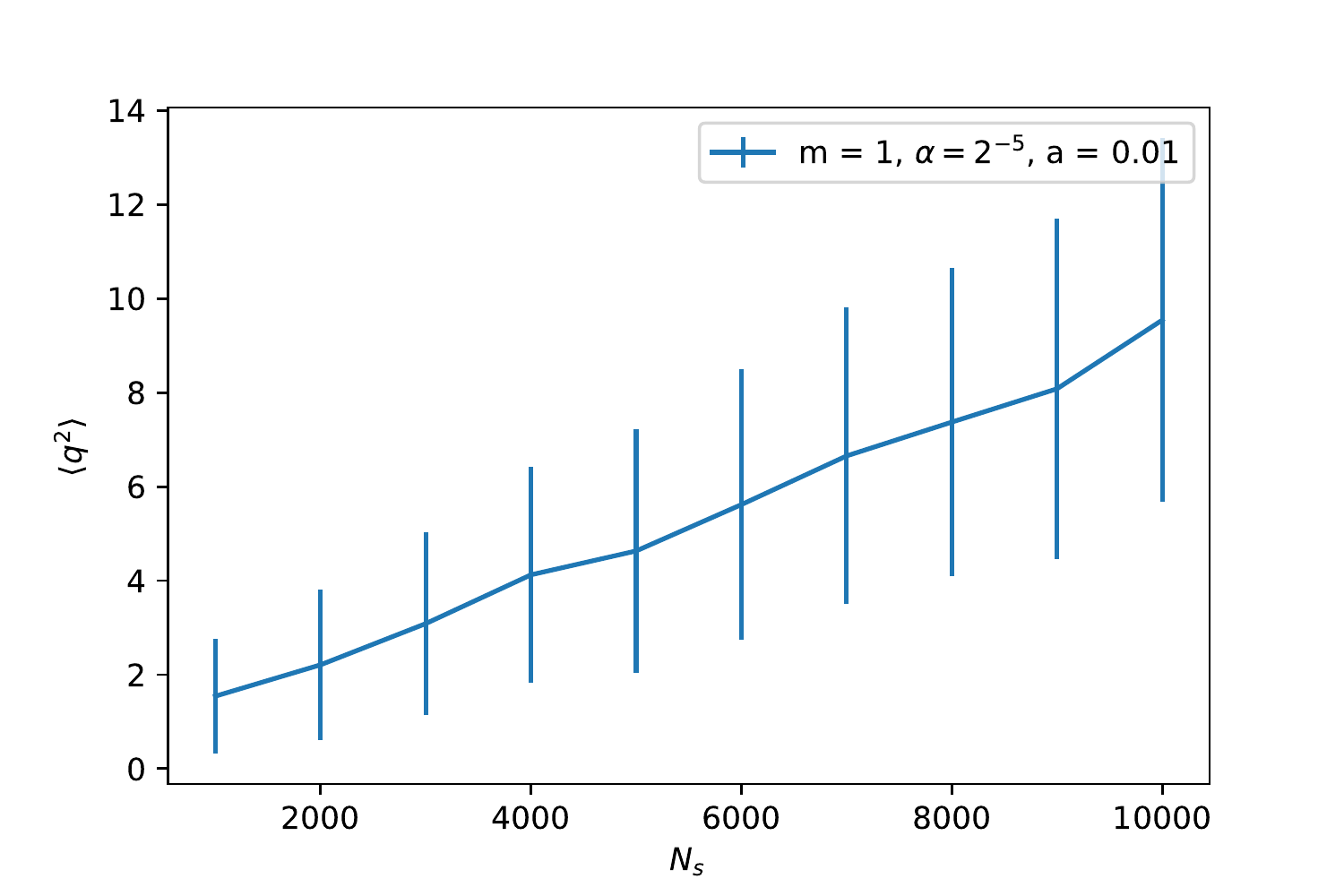}
    \caption{Dependence of $\langle q ^ 2 \rangle$ on $N _ s$ at $m = 1$, $a = 0.01$ and different values of $N _ s$.}
\end{figure}

\subsection{Critical coupling}
The calculations of critical coupling demonstrates the following dependence on mass (see Figure 4). Systems with small mass values can be considered in  ultra-relativistic (UR) regime, which is characterized by no mass dependency
\begin{equation}
    H _ {UR} = |p| - \frac{\alpha e ^ 2}{\sqrt{q ^ 2 + a ^ 2}}.
\end{equation}
Critical coupling constant decreases with mass in medium mass range. Huge values of mass correspond to weak dependence of coupling constant on mass. 
\begin{figure}[ht]
    \centering
    \includegraphics[width=0.75\linewidth]{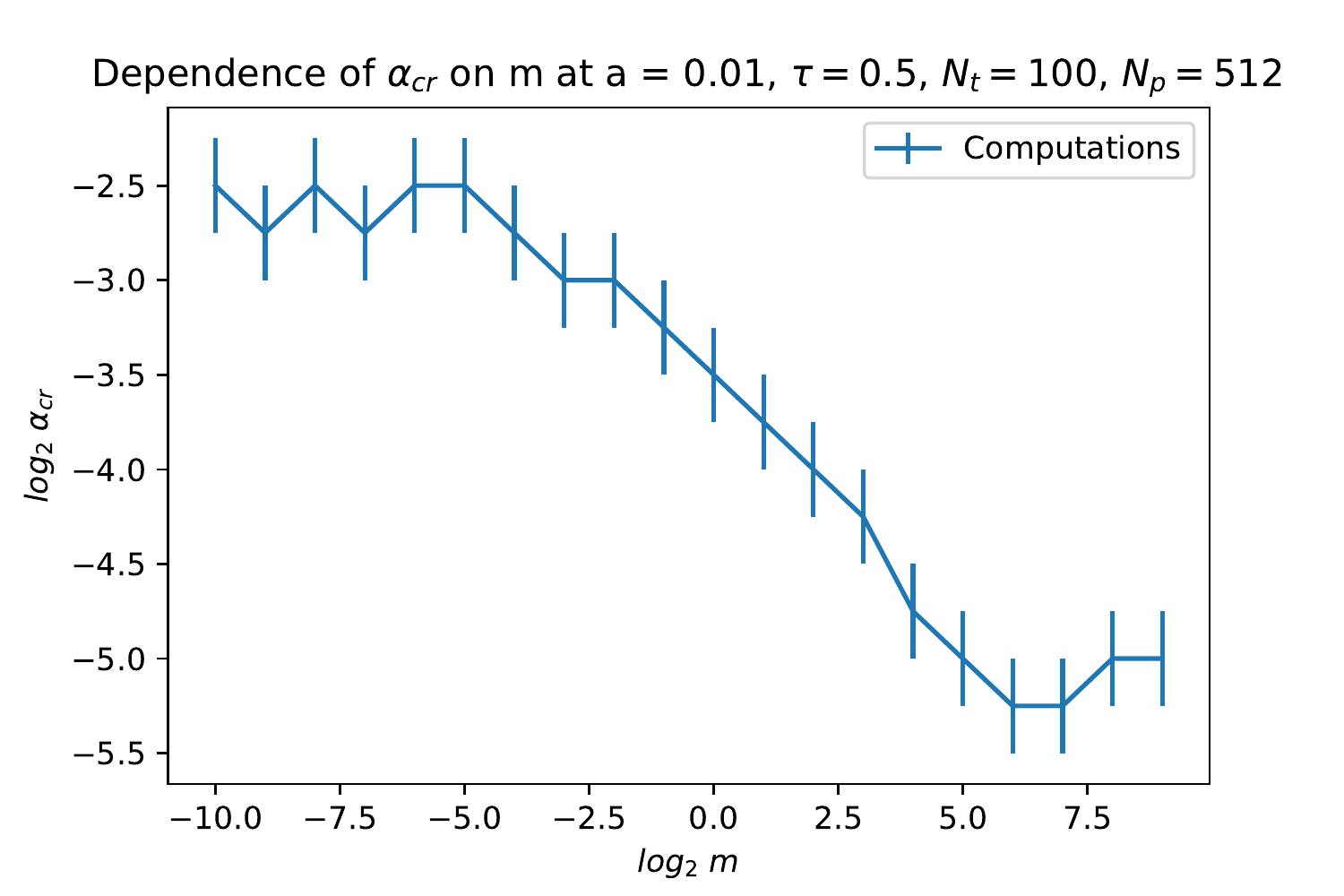}
    \caption{Dependence of critical coupling constant $\alpha_{cr}$ on the mass 
    of the particle  $m$, $a = 0.01$.}
\end{figure}

\section{Conclusion}

In our work the relativistic 1D Coulomb problem was studied by using Relativistic Path Integral Monte Carlo method. 
The interesting feature  of this relativistic generalization of PIMC method is the fact that the statistical weights of the trajectories do not correspond to classical action like in Feynman-Kac formula but have the spatial form. In our previous study this approach has been tested on the relativistic oscillator problem \cite{inp}. Now we use the Relativistic PIMC for the investigation of the relativistic 1D Coulomb problem. In contrast to relativistic oscillator problem, relativistic 1D Coulomb problem deals with the singular potential and has no analytical solution. Fortunately, there is virial theorem which we can use for the testing of non-relativistic regime in this problem. In addition, the virial theorem helps us to find the values of the model parameters for relativistic and non-relativistic regimes. As expected, the relativistic regime appears at small masses of the particle and (or) at strong coupling. Another interesting result was obtained for critical coupling at which the bound state disappears.  

\section*{Acknowledgements}

A.I.  acknowledge the support from BASIS Foundation for Development of Theoretical Physics and Mathematics (grant number 17-21-103-1). The work of O.P., which consisted physical interpretation of Monte Carlo data, was supported by grant
from the Russian Science Foundation (project number 16-12-10059).

 \end{document}